\documentclass[submission,copyright]{eptcs}
\providecommand{\event}{Wivace 2013}
\usepackage{breakurl}
\usepackage{amsmath}
\usepackage{graphicx}
\usepackage{subfig}
\usepackage{amssymb}

\newcommand{\XX}{\mathbf{X}}

\title{GPU-powered Simulation Methodologies\\ for Biological Systems}

\author{\small Daniela Besozzi$^a$ \quad Giulio Caravagna$^b$ $\quad$ Paolo Cazzaniga$^c$ $\quad$\\
\small Marco Nobile$^b$ \quad Dario Pescini$^d$ \quad Alessandro Re$^b$ \\
\institute{$^a$Universit\`{a} degli Studi di Milano, Dipartimento di Informatica\\ Via Comelico 39, 20135 Milano, Italy}
\email{besozzi@di.unimi.it}
\institute{$^b$Universit\`{a} degli Studi di Milano-Bicocca, Dipartimento di Informatica, Sistemistica e Comunicazione\\ Viale Sarca 336, 20126 Milano, Italy}
\email{giulio.caravagna/nobile@disco.unimib.it, a.re4@campus.unimib.it}
\institute{$^c$Universit\`{a} degli Studi di Bergamo, Dipartimento di Scienze Umane e Sociali\\ Piazzale S. Agostino 2, 24129 Bergamo, Italy}
\email{paolo.cazzaniga@unibg.it}
\institute{$^d$Universit\`{a} degli Studi di Milano-Bicocca, Dipartimento di Statistica e Metodi Quantitativi\\ Via Bicocca degli Arcimboldi 8, 20126 Milano, Italy}
\email{dario.pescini@unimib.it}
}

\def\titlerunning{GPU-powered Simulation Methodologies for Biological Systems}
\def\authorrunning{D. Besozzi et al.}
\begin{document}
\maketitle

\vspace{-0.2cm}
\section*{Aims and motivation}

In the last decades, the study of biological systems witnessed a pervasive cross-fertilization between experimental investigation and computational methods, thanks to the convergence of the so-called high-throughput era with the easy access to high performance computing resources.
This combination gave rise to the development of new methodologies, able to tackle the complexity of biological systems in a \emph{quantitative} manner.
For instance, given a mathematical formalization of complex biological networks, it is nowadays possible to determine their emergent dynamical and structural properties with whole-system based approaches, in order to make predictions on the way these systems behave in normal conditions or how they react to different perturbations \cite{book_system_modelling}.

In this context -- according to the system under investigation, to the experimental data at hand and to the biological questions one is expected to unravel -- the choice of the most suitable modeling approach is fundamental to properly elucidate the underlying physics of the target system.
Standard modeling and simulation approaches span from the stochastic to the deterministic formalization of chemical kinetics \cite{Wilkinson2009}, also considering the possibility to combine these two approaches into hybrid methods \cite{Kurtz,Davis1984,salis2005}.
The conditions of applicability of each method are strictly related to the characteristics of the biological system under investigation, as briefly presented in the next section.

Given either a stochastic, deterministic or hybrid mathematical model, computer algorithms can then be exploited to investigate the temporal evolution of the corresponding biological system: starting from different initial conditions of the model (given in terms of distinct values for the kinetic constants, the initial molecular amounts, etc.), different emergent behaviors of the system can be determined.
This fact highlights the relevance of setting a good parameterization for the model; as a consequence, the exploration of high-dimensional parameter spaces will allow to investigate the system functioning across a wide spectrum of natural conditions, as well as to derive statistically meaningful properties.
These issues play a fundamental role in the computational analysis of biological systems, which usually exploit methodologies based on parameter sweep analysis, sensitivity analysis, parameter estimation and reverse engineering of model topologies (see \cite{Aldridge2006,Chou09} and references therein).

All these methods rely on the repetition of a large number of simulations, therefore demanding the reduction of the computational costs for everyday applicability.
An emergent technology suitable to tame this problem is the General-Purpose Graphics Processing Unit (GPGPU) paradigm, in which the parallel computation capabilities of modern video cards are exploited for general purpose computations.
Indeed, GPGPU allows to perform multiple analysis in parallel, using cheap, diffused and highly efficient multi-core devices.
Despite the remarkable advantages in terms of the achievable computational speedup, computing with GPUs requires specific programming skill, since GPU-based programming substantially differs from CPU-based
computing; as a consequence, scientific applications on GPUs risk to remain a niche for few specialists.
To avoid such limitations, we are developing GPU-powered simulation algorithms for stochastic, deterministic and hybrid modeling approaches, so that also users with no knowledge of GPUs hardware and programming can easily access the computing power of graphics engines.

\vspace{-0.2cm}
\section*{Methods}

To investigate the behavior of a given biological system, the choice of the most adequate modeling and simulation approach can be carried out, in general, according to the molecular amounts of the chemical species occurring in the system, or even to the temporal scale of the biological phenomena that one wishes to reproduce and analyze.

When the biological system is characterized by chemicals species occurring in small molecular amounts (in the order of units, tens, or a few hundreds of molecules), accounting for the {\em intrinsic random  fluctuations}  is fundamental to correctly mimic the behavior of the target system \cite{Eldar_noise_Nature2010}.
In this case, stochastic approaches should be adopted.
Under the assumption of spatial homogeneity and thermal equilibrium, the system states can be formally described by a continuous-time discrete-value  random variable  $\XX(t)$, that denotes the number of molecules of each species.
Changes in the system state are determined by firing chemical reactions, which describe the interactions between the chemical species occurring in the system.
Understanding the system dynamics therefore reduces to estimating the probability of $\XX(t)$ to be in each possible (chemical) state in time, given some initial condition.
This probability changes according to the  {\em Chemical Master Equation}, which is usually unfeasible, especially for large systems, due to the unpractical numerical calculations required to solve it.
However, sample paths of $\XX(t)$  and  numerical estimates of the probability of any chemical state can be obtained algorithmically, whereby $\XX(t)$ is a {\em Continuous-Time Markov Chain}, whose states are the values of $\XX(t)$ \cite{Gill77}.
To speedup the simulation  of $\XX(t)$,  approximation techniques  either based on the  {\em Chemical Langevin Equation} \cite{Cao06}  or on quasi-steady-state assumptions  \cite{ssSSA} are often used.
Coupled intrinsic/extrinsic environmental fluctuations can also be considered \cite{GiulioPlos}.

When most of the chemical species are present in the system in large amounts (in the order of thousands of molecules or more), the intrinsic fluctuations of $\XX(t)$ get averaged out, and $\XX(t)$ is well approximated by a set of {\em Ordinary Differential Equations}, often termed  {\em Reaction-Rate Equations}.
Generally, the unique path of $\XX(t)$ is numerically evaluated via numerical integration algorithms.
It was shown that, when the system is spatially uniform and at the thermodynamic limit, the stochastic  and the mean-field model have the same behavior \cite{Kurtz}.

When a clear separation between the temporal scales of the reactions exists, hybrid -- stochastic and deterministic -- approaches can be adopted.
This is the case when, for instance, kinetic constants or molecular amounts are separated by several orders of magnitude, which is likely to happen in multiscale modeling of cellular and intercellular dynamics \cite{salis2005}.
Hybrid approaches allow to account for intrinsic random fluctuations due to low-amount species, at the same time integrating an appropriate mean-field approximation concerning the abundant chemicals \cite{GiulioJTB,GiulioBMC}.
Technically, here $\XX(t)$ describes a {\em Piecewise Deterministic Markov Process} \cite{Davis1984}.

Whatever the mathematical representation of the model, the simulations of the system dynamics in a given set of initial conditions -- each one characterized by the same or a different parameterization with respect to the other conditions -- are generally executed in a serial fashion on standard CPU personal computers, thus causing high computational costs.
However, all these simulations are mutually independent, therefore the computational burden can be strongly reduced by exploiting a parallel architecture.
Among the existing solutions, we are considering GPGPU computing for the implementation of the aforementioned simulation algorithms, in order to gain a consistent reduction in the overall computational time required to fully analyze a biological model \cite{Nobile2013,tauGPU}.
In particular, we are adopting nVidia's CUDA, a GPGPU library that combines the single instruction multiple data architecture and multi-threading.
Using CUDA's naming conventions, the programmer implements a kernel loaded from the host (the CPU) to the devices (one or more GPUs), replicated in many copies named \emph{threads} (where each thread corresponds to a dynamics simulation).
Threads are logically organized in structures named \emph{blocks} which, in turn, are organized in \emph{grids}.
Threads are executed in groups named \emph{warps} (corresponding to 16 threads), whose scheduling is handled by the driver which dispatches the work on the multiple streaming multiprocessors, thus allowing a transparent scaling of performances on different GPUs.

\vspace{-0.2cm}

\section*{Discussion}

In living cells, many processes are regulated by feedback mechanisms, which are usually interlaced in complex regulatory networks and can overall function to either attenuate or amplify molecular noise and stochasticity.
In this context, a clear example of the need of \emph{ad hoc} tools to properly analyze the functioning of biological systems when different temporal scales and molecular amounts simultaneously occur, is the Ras/cAMP/PKA pathway in yeast.
This signalling pathway regulates cell metabolism and cell cycle progression in response to nutritional sensing and stress conditions through multiple feedback loops \cite{thev94}; because of such complex interplay, it is not easy to predict its behavior in different growth conditions or in response to various stress signals.
To understand the role of the negative feedback controls that are present in this system, in \cite{besozziEURASIP,Pescini2011,Cazzaniga2008} we defined and analyzed a mathematical model of the Ras/cAMP/PKA pathway, focusing our attention on the mechanisms that allow the emergence of oscillatory regimes in cAMP dynamics.

The model of the Ras/cAMP/PKA pathway was initially developed according to the stochastic formulation of chemical kinetics \cite{Gill77}, and then translated into a ``generalized mass-action based'' model.
The comparison between stochastic and deterministic simulations evidenced different quantitative and qualitative results under different conditions \cite{besozziEURASIP}, suggesting a functional role of intrinsic noise.
The application of hybrid approach on this model is currently in progress, in order to accurately deal with the small molecular amounts of some pivotal proteins (Cdc25, Ira2) and the large amount of guanine nucleotides (GTP, GDP), which both have shown to represent key regulatory factors for the establishment of oscillatory regimes within the pathway.

We have exploited our GPU implementations of stochastic, deterministic and hybrid simulation algorithms to execute the massive number of simulations that were necessary to carry out an extensive analysis of this signaling pathway.
Our preliminary tests confirmed that the GPU implementation is from 10 to 100 times faster than the standard CPU implementation of the same simulation algorithms, therefore allowing deeper investigations of the system.
The general GPU suite that we are developing is therefore a promising tool to easily perform intensive and fast computational analysis of biological systems.

\bibliographystyle{eptcs}
\bibliography{wivace}

\end{document}